\documentclass[twocolumn,showpacs,preprintnumbers,amsmath,amssymb]{revtex4}
\usepackage{amsfonts}% Physical Review B
\usepackage{enumitem}
\usepackage{epsf}
\usepackage{anysize}
\marginsize{1 in}{1 in}{1in}{1.40in}

\usepackage{graphicx}% Include figure files
\usepackage{subfigure}
\usepackage{dcolumn}% Align table columns on decimal point
\usepackage{bm}% bold math
\usepackage{amsmath}

\usepackage{amssymb}
\usepackage{mathrsfs}
\usepackage[latin1]{inputenc}
\usepackage{pstricks,pst-node,pst-text,pst-3d}

\usepackage[all]{xy}

\usepackage{setspace}
\usepackage{tikz}
\usetikzlibrary{shapes,arrows}
\usepackage{lipsum}
\usepackage{lmodern}
\usepackage{tcolorbox}

\usepackage{diagbox}
\usepackage[ ruled ]{algorithm2e}
\usepackage{algpseudocode}
\usepackage{textcomp}
\usepackage{graphicx}
\usepackage{subfigure}

\begin{document}

\title{ Risk-sensitive Optimization for Robust Quantum Controls}
\author{Xiaozhen Ge}
\affiliation{Department of Automation, Tsinghua University, Beijing 100084, China}
\author{Re-Bing Wu}
\email{rbwu@tsinghua.edu.cn}
\affiliation{Department of Automation, Tsinghua University, Beijing 100084, China}

\begin{abstract}

Highly accurate and robust control of quantum operations is vital for the realization of error-correctible quantum computation. In this paper, we show that the robustness of high-precision controls can be remarkably enhanced through sampling-based stochastic optimization of a risk-sensitive loss function. Following the stochastic gradient-descent direction of this loss function, the optimization is guided to penalize poor-performance uncertainty samples in a tunable manner. We propose two algorithms, which are termed as the risk-sensitive GRAPE and the adaptive risk-sensitive GRAPE. Their effectiveness is demonstrated by numerical simulations, which is shown to be able to achieve high control robustness while maintaining high fidelity. 
\end{abstract}

\maketitle

\section{introduction}
In quantum computation, extremely high-precision control of quantum operations is highly demanded~\cite{Nielsen2011Quantum,Surface,2019Quantum}. In addition to the precision, the robustness is also required against uncertainties and noises in realistic quantum devices, e.g., the pulse distortion, the dephasing noise and crosstalks~\cite{2019Iterative}. In the literature, various protocols have been proposed to improve robustness, including the combination of slow-varying pulses and stimulated Raman adiabatic passage (STIRAP) that is insensitive to pulse shape errors~\cite{torosov2013composite,vitanov2017stimulated}, the fast sequences of periodic pulses for dynamically decoupling decoherence noises~\cite{Viola1999Dynamical,genov2017arbitrarily, sekiguchi2019dynamical}, and the derivative removal adiabatic gates (DRAG) method for eliminating the leakage to higher levels~\cite{2009Simple,2010Optimized,2016Efficient}. 

Recently, gradient-based optimization is introduced to systematically train robust control pulses against uncertainties and noises. The basic idea thereof is to formulate the robust control design as the minimization of an empirical loss function evaluated over uncertainty samples. The loss function is usually chosen as the average error or the worst-case error, based on which various gradient-based algorithms were proposed for the training of robust controls, e.g., the sampling-based learning~\cite{Chen2014Sampling,Dong2015Sampling}, stochastic gradient-based algorithms~\cite{Wu2018Deep,Gabriel}, the sequential convex programming (SCP)~\cite{kosut2013robust,QAOA} and the adversarial training based a-GRAPE~\cite{AGRAPE}. These algorithms have been shown effective in improving the robustness against various uncertainties, e.g., coupling uncertainty~\cite{Wu2018Deep,AGRAPE}, energy broadening~\cite{Zhang2014Robust,Chen2014Sampling,Dong2015Sampling}, inhomogeneity of control field pulses~\cite{Chen2014Sampling,Dong2015Sampling,turinici2019stochastic} and clock noises~\cite{ding2019robust}.

\begin{figure}\label{pdf}
	\begin{center}
		\includegraphics[width=1\columnwidth]{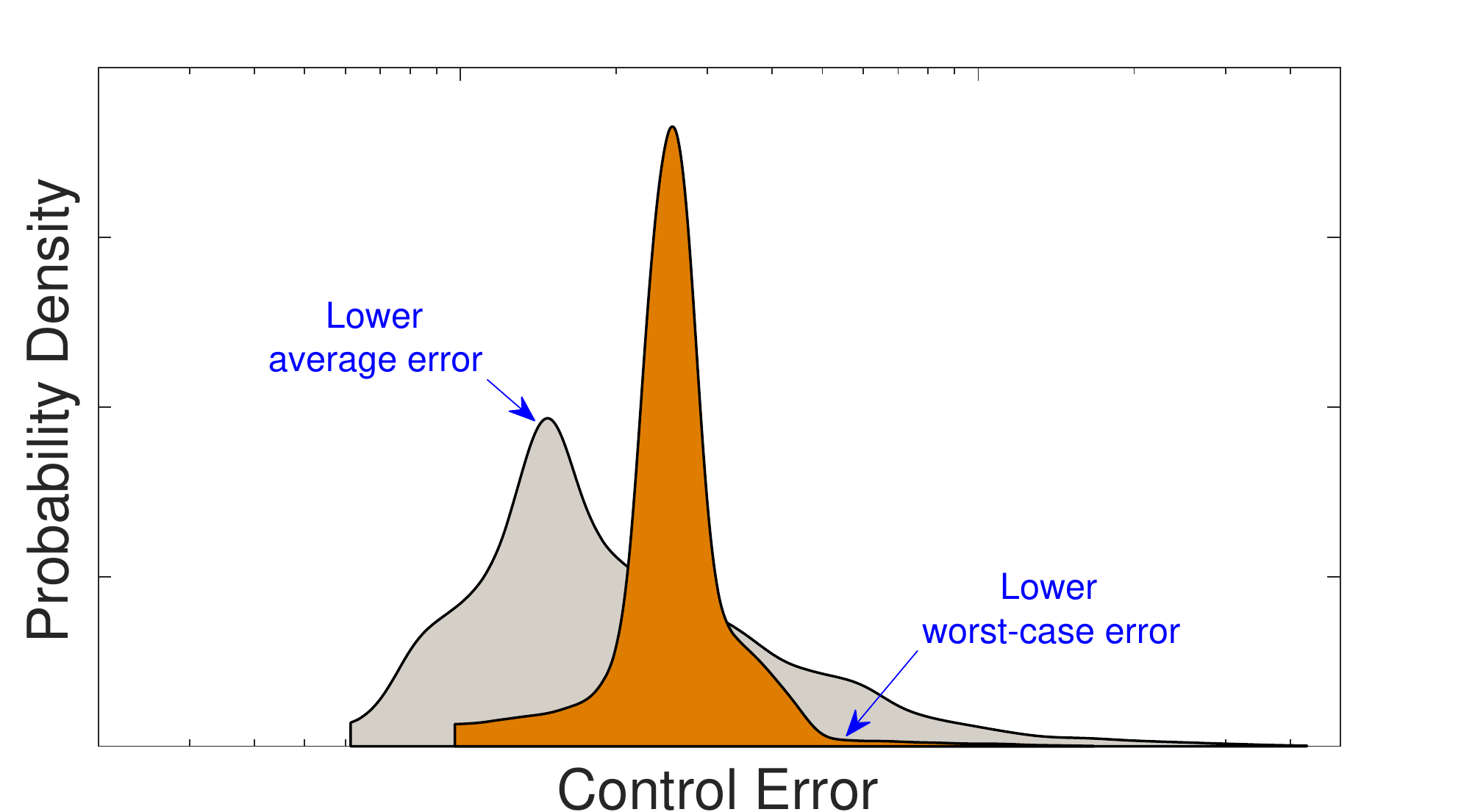}
	\end{center}
	\caption{The comparison of robustness between controls optimized with average error and worst-case error. The envelopes of the gray and orange regions respectively correspond to the probability density function (pdf) of the error under the designed controls with respect to the average error and the worst-case error.}
	
\end{figure}

The choice of the loss function is central to the design of robust optimization algorithms. The average error is often the first choice because it is easier to evaluate and thus to optimize. However, as is schematically shown in Fig.~\ref{pdf}, the resulting controls may have poor worst-case performance due to the lack of a control over the variance of the error. The worst-case error based optimization can effectively broaden the regimes of robustness, but the performance in the high-precision regime may not be satisfactory. These facts show that the two loss functions make different trade-offs between the precision and the robustness. In our recent work~\cite{AGRAPE}, we illustrate that the a-GRAPE approaches can adjust the trade-off by purposely using poor-performance uncertainty samples. However, many hyper-parameters in the algorithm have to be empirically tuned, which is computationally costly when the control parameters or the uncertainty parameters are high-dimensional.

To alleviate the conflict between the desired precision and the high robustness, we propose in this paper the risk-sensitive (RS) optimization approach which is illuminated by classical control theory~\cite{James1994Risk,Fleming1995Risk,Dupuis1997Risk,Dupuis2000Robust,917658,dupuis1998robust}. The training is made sensitive to the poor-performance uncertainty samples, which takes both the advantages of the cases subject to the average error and the worst-case error. In the literature, the RS criterion has been considered in the design of feedback controls for linear quantum systems~\cite{James2004,Helon2007Quantum}. However, we have not seen any studies on the design of robust open-loop controls, and this motivates our studies that lead to RS-based robust quantum control design algorithms. 

The rest of this paper is structured as follows. Section \ref{sec2} formulates the robust quantum control problem with respect to the RS criterion and presents the RS-based GRAPE algorithms for the training of robust quantum controls. In Sec.~\ref{simulation}, the effectiveness of the proposed algorithms is illustrated through numerical simulations. Finally, the conclusion is made in Sec.~\ref{discussion}.

\section{ The risk-sensitive optimization}\label{sec2}

Consider an $N$-dimensional closed quantum system that is described by the following controlled Schr\"{o}dinger equation:
\begin{equation}
 \frac{{\rm d}}{{\rm d}t}U(t) = -iH\left[t;\boldsymbol{u},\boldsymbol{\epsilon}\right]U(t).
\end{equation}
Here, $U(\cdot)\in \mathbb{C}^{N\times N}$ represents the unitary propagator starting from $U(0)=\mathbb{I}_{N}$, where $\mathbb{I}_N$ is the identity matrix. The time-varying Hamiltonian $H\left[t;\boldsymbol{u},\boldsymbol{ \epsilon}\right]$ depends on a vector of control parameters $\boldsymbol{u}$ (e.g., in-phase and quadrature amplitudes that vary in time, or phases and amplitudes of laser pulses in frequency-domain) and a random vector of uncertainty parameters $\boldsymbol{\epsilon}$ (e.g., environmental noises, or imprecisely identified parameters) following some probability distribution $P(\boldsymbol{\epsilon})$. The robust control aims at achieving a desired gate operation $U_f$ at some final time $T$ with high accuracy, which must be also insensitive to the uncertainties so as to maintain high performance no matter how $\boldsymbol{\epsilon}$ varies.

The design of robust controls needs to be based on proper measure of the robustness. Let
\begin{equation} L(\boldsymbol{u},\boldsymbol{\epsilon})=N^{-2}\|U(T;\boldsymbol{u},\boldsymbol{\epsilon})-U_f\|^2
\end{equation}
be the gate infidelity (or error) under the control $\boldsymbol{u}$ and the uncertainty $\boldsymbol{\epsilon}$. If the probability distribution $P(\boldsymbol{\epsilon})$ of $\boldsymbol{\epsilon}$ is \textit{a priori} known, we can use the average infidelity
\begin{equation*}  F(\boldsymbol{u})=\int_{\boldsymbol{\epsilon}}{\rm d}P(\boldsymbol{\epsilon})L(\boldsymbol{u},\boldsymbol{\epsilon}). 
\end{equation*}
Alternatively, we can use the worst-case infidelity
 \begin{equation*}
 F(\boldsymbol{u})=\max_{\boldsymbol{\epsilon}}L(\boldsymbol{u},\boldsymbol{\epsilon})
 \end{equation*} 
 that is irrelevant with the distribution $P(\boldsymbol{\epsilon})$. To make better use of both measures, we consider the following risk-sensitive (RS) criterion
\begin{equation}\label{rs}
F_\mu(\boldsymbol{u})
=\int_{\boldsymbol{\epsilon}}{\rm d}P(\boldsymbol{\epsilon})V_{\mu}\left[L(\boldsymbol{u},\boldsymbol{\epsilon})\right],
\end{equation}
where $V_{\mu}[\cdot]: \mathbb{R}^+\rightarrow \mathbb{R}^+$ is a pre-selected differentiable utility function parameterized by the sensitivity parameter $\mu$. It is introduced to put different weights on uncertainties according to the resulting errors. Specifically, the objective function (\ref{rs}) is risk-averse if the uncertainty samples with poorer performance are assigned to greater weight, which is the case we consider in this paper. The role of the parameter $\mu$ is designed to tune the degree of bias in the evaluation of control errors that emphasizes on large-error uncertainty samples. For example, the utility can be chosen as the exponential form $V_\mu=e^{\mu L}$ (or the HARA utility $V_\mu=L^\mu$), which approaches the average infidelity (i.e., the risk-neutral case) when $\mu\rightarrow 0^+$ (or $\mu\rightarrow 1$), and equivalent to the worst-case infidelity when $\mu\rightarrow +\infty$. Therefore, the RS loss function can be taken as the combination of average infidelity and the worst-case infidelity, and their trade-off is balanced by $\mu$. 

As described above, under an appropriately selected utility function $V_\mu[\cdot]$, the RS loss function is more aggressive than the average infidelity but less conservative than the worst-case infidelity, which makes it possible to take advantages of both performance indices. Meanwhile, it is easy to be optimized just like the average infidelity, and many effective gradient-based optimization methods that have matured in machine learning can be directly applied.

\begin{algorithm}[htb]\label{alg1}
	\caption{ RS GRAPE }
	
	\noindent{\bfseries Initialize:}
	
	\setlength\parindent{1em} an initial guess on the control $\boldsymbol{u}$;

	\setlength\parindent{1em} the mini-batch size $M$; 
	
	\setlength\parindent{1em} the sensitivity parameter $\mu$;
	
	\noindent{\bfseries Repeat:}
	\begin{itemize}[itemsep=1pt, topsep=2pt,partopsep=2pt,parsep=1pt]
		\item[(1)] Randomly select a mini-batch of uncertainty samples $B=\{\boldsymbol{\epsilon}_i  \}_{i=1}^M$;
		\item[(2)] Compute the gradient $\boldsymbol{g}_\mu[\boldsymbol{u}, B]$ by using the sampled uncertainty mini-batch, and update the controls by gradient-descent methods, e.g.,
		{\setlength\abovedisplayskip{1pt}
			\setlength\belowdisplayskip{1pt}
			\begin{equation*} 
			\boldsymbol{u}\leftarrow \boldsymbol{u}-\alpha\cdot\boldsymbol{g}_\mu[\boldsymbol{u}, B],
			\end{equation*}}
		where $\alpha$ denotes the learning rate.
	\end{itemize}
	
	\noindent {\bfseries Until} Stopping criteria satisfied.
\end{algorithm}

\subsection{Stochastic gradient-descent algorithms with fixed sensitivity}

The simplest way of using the RS function is to fix the sensitivity parameter, where the corresponding training process of the robust controls is described as in Algorithm~\ref{alg1}. Let $\boldsymbol{u}^{(k)}$ be the optimized control in the $k$-th iteration, and $B^{(k)}=\{\boldsymbol{\epsilon}^{(k)}_1, \boldsymbol{\epsilon}^{(k)}_2,\cdots, \boldsymbol{\epsilon}^{(k)}_M\}$ is a mini-batch of $M$ uncertainty samples that are randomly generated according to the probability distribution $P(\boldsymbol{\epsilon})$. Then, the empirical RS loss can be approximated by 
\begin{equation*}
\hat{F}_{\mu}(\boldsymbol{u}^{(k)},B^{(k)})=\frac{1}{M}\sum_{i=1}^{M} V_{\mu}\left[L(\boldsymbol{u}^{(k)},\boldsymbol{\epsilon}_i^{(k)})\right].
\end{equation*}
It is straightforward to derive the gradient formula
\begin{eqnarray}\label{g1}
\boldsymbol{g}_\mu(\boldsymbol{u}^{(k)},B^{(k)})
&=&\frac{1}{M}\sum_{i=1}^M %\frac{{\rm d}
 V'_{\mu}\left[L(\boldsymbol{u}^{(k)},\boldsymbol{\epsilon}_i^{(k)})\right]%} {{\rm d} L}
\cdot\nonumber\\
&&\frac{\delta }{\delta \boldsymbol{u}}L(\boldsymbol{u}^{(k)},\boldsymbol{\epsilon}_i^{(k)}),
\end{eqnarray}
where $V'_\mu:={\rm d}V_\mu/{\rm d} L$ denotes the derivative of the utility function with respect to the gate infidelity $L$, and the variation $\delta L/\delta\boldsymbol{u}$ can be evaluated through the formula derived in the GRAPE method~\cite{optimal}. Normalizing the weights assigned to the directions, we can rewrite the gradient as
\begin{equation}\label{g2}
\boldsymbol{g}_\mu[\boldsymbol{u}^{(k)},B^{(k)}]=\sum_{i=1}^{M} \omega_i^{(k)} \frac{\delta}{\delta \boldsymbol{u}}L(\boldsymbol{u}^{(k)},\boldsymbol{\epsilon}_i^{(k)})
\end{equation}
with the $\mu$-dependent weight
\begin{equation}\label{weight}
\omega_i^{(k)}=\frac{V_{\mu}'[L(\boldsymbol{u}^{(k)},\boldsymbol{ \epsilon}_i^{(k)})]}{\sum_{j=1}^M V_{\mu}'[L(\boldsymbol{u}^{(k)},\boldsymbol{ \epsilon}_j^{(k)})]}.
\end{equation}
Typically, we have
\begin{equation}
\omega_i^{(k)}=\frac{e^{\mu L(\boldsymbol{u}^{(k)},\boldsymbol{ \epsilon}_i^{(k)})}}{\sum_{j=1}^M e^{\mu L(\boldsymbol{u}^{(k)},\boldsymbol{ \epsilon}_j^{(k)})}}
\end{equation}
when $V_\mu[L]= e^{\mu L}$, and 
 \begin{equation}
 \omega_i^{(k)}= \frac{[L(\boldsymbol{u}^{(k)},\boldsymbol{ \epsilon}_i^{(k)})]^{\mu-1}}{\sum_{j=1}^M  [L(\boldsymbol{u}^{(k)},\boldsymbol{ \epsilon}_j^{(k)})]^{\mu-1}}
 \end{equation}
 when $V_\mu[L]=L^{\mu}$. Along the obtained stochastic gradient (\ref{g2}) evaluated by the mini-batch of uncertainty samples, we can update the control using stochastic optimization algorithms (e.g., the Adam method~\cite{Goodfellow2016Deep}).

It is clear that the risk-sensitive gradient (\ref{g2}) re-evaluates the importance of each uncertainty sample by the utility function $V_\mu$. When $V_\mu'[L]> 0$ and ${\rm d}V_\mu'[L]/{\rm d}{L}>0$ (e.g., the exponential utility for $\mu>0$ and the HARA utility for $\mu>1$), it weights heavier on uncertainty samples with larger control errors.

\subsection{Stochastic gradient-descent algorithms with adaptive sensitivity}

  In the above proposed algorithm, the risk-averse training automatically penalizes ``bad" uncertainty samples, and in this way the control robustness is supposed to be enhanced when the sampled uncertainties have diverse performance. However, the performance diversity tends to vanish when the control is gradually hardened, and eventually the risk-sensitive training will become ineffective. Under such circumstance, the sensitivity parameter $\mu$ should be adaptively tuned to amplifying the diversity so that the poorer-performance samples can be adequately addressed.

In Algorithm~\ref{alg}, we propose a strategy for adaptively tuning $\mu$ according to diversity of uncertainty samples. Denote by $r(\mu):=\max_{1\leq i\leq M}\omega_i$ the diversity degree (i.e., the weight assigned to the worst sample). Given a desired diversity degree $r^*$, we choose the parameter $\mu$ such that $r(\mu)=r^*$ in each iteration. In this way, the worst sample is always adequately penalized even if the difference between sample performances is very small.

In practical applications, the desired diversity degree $r^*$ can be empirically selected, and $\mu$ is obtained by numerically solving the nonlinear equation $r(\mu)=r^*$. It is not hard to prove that for the exponential utility and the HARA utility, the solution always exists for an arbitary $r^*\in[1/M,1]$.

\begin{algorithm}[htb]\label{alg}
		\caption{Adaptive RS GRAPE     }
		
		\noindent{\bfseries Initialize:}
		
		\setlength\parindent{1em} an initial guess on the control $\boldsymbol{u}$;	
		
		\setlength\parindent{1em} the mini-batch size $M$; 
		
		\setlength\parindent{1em} the diversity degree $r^*\in[1/M, 1]$;
		
		\noindent{\bfseries Repeat:}
		\begin{itemize}[itemsep=1pt, topsep=2pt,partopsep=2pt,parsep=1pt]
			\item[(1)] Randomly select a mini-batch of uncertainty samples $B=\{\boldsymbol{\epsilon}_i  \}_{i=1}^M$;
			\item[(2)] Compute the gradients $\delta L(\boldsymbol{u},\boldsymbol{\epsilon}_i)/\delta \boldsymbol{u}$ for each sample as in \cite{khaneja2005optimal}, and determine the parameter $\mu$ as the solution of the following equation
			 {\setlength\abovedisplayskip{1pt}
				\setlength\belowdisplayskip{1pt}
				\begin{equation*} \label{mu}
				\max_i \omega_i(\mu)=r^*;
				\end{equation*}}
			\item[(3)] Calculate the weight $\omega_i$  and the weighted gradient $\boldsymbol{g}_\mu[\boldsymbol{u},B]$. Then, update the controls by gradient-descent methods, e.g.,
			{\setlength\abovedisplayskip{1pt}
				\setlength\belowdisplayskip{1pt}
				\begin{equation*} 
			\boldsymbol{u}\leftarrow \boldsymbol{u}-\alpha\cdot\boldsymbol{g}_\mu[\boldsymbol{u}, B].
				\end{equation*}}
		\end{itemize}

		\noindent {\bfseries Until} Stopping criteria satisfied.
	\end{algorithm}

\section{Simulation Examples}\label{simulation}
In this section, we numerically test the effectiveness of our proposed RS GRAPE and adaptive RS GRAPE approaches for the design of robust quantum controls. We consider a three-qubit control system, whose Hamiltonian is as follows:
\begin{eqnarray*}
	H(t)&=&J_{12}(1+\epsilon_1)\sigma_{1}^z\sigma_{2}^z+J_{23}(1+\epsilon_2)\sigma_{2}^z\sigma_{3}^z\\
	&&+\sum_{k=1}^3\left[u_{kx}(t)\sigma_{k}^x+u_{ky}(t)\sigma_{k}^y\right],
\end{eqnarray*}
where $\sigma_k^{x,y,z}$ denotes the Pauli operator associated with the $k$-th qubit, and $J_{12}=J_{23}=10$ MHz are the identified qubit-qubit coupling strength. The uncertainty parameters $\epsilon_1$ and $\epsilon_2$ represent the identification errors that are uniformly distributed over $[-0.2,0.2]$. The control pulses $u_{kx}(t)$ and $u_{ky}(t)$ are delivered to the $k$-th qubit along the $x$-axis and $y$-axis, respectively, which are evenly divided into $100$ piecewise-constant sub-pulses over the time interval $[0,1]$ $\mu s$. The target unitary operation is selected as the Toffoli gate.

In the simulations, the batchsize is selected as $M=10$, and the initial controls are chosen to be $u_{kx}(t)=A_k\sin(\omega_k t+\phi_k)$ and $u_{ky}(t)=A_k\cos(\omega_k t +\phi_k)$, where $A_k$, $\omega_k$ and $\phi_k$ are randomly generated. Moreover, we use the Adam method to iteratively update the learning rate $\alpha$, which has been broadly applied for training deep neural networks~\cite{Adam,Goodfellow2016Deep}.

\begin{figure*}
	\begin{center}
		\includegraphics[width=0.9\linewidth]{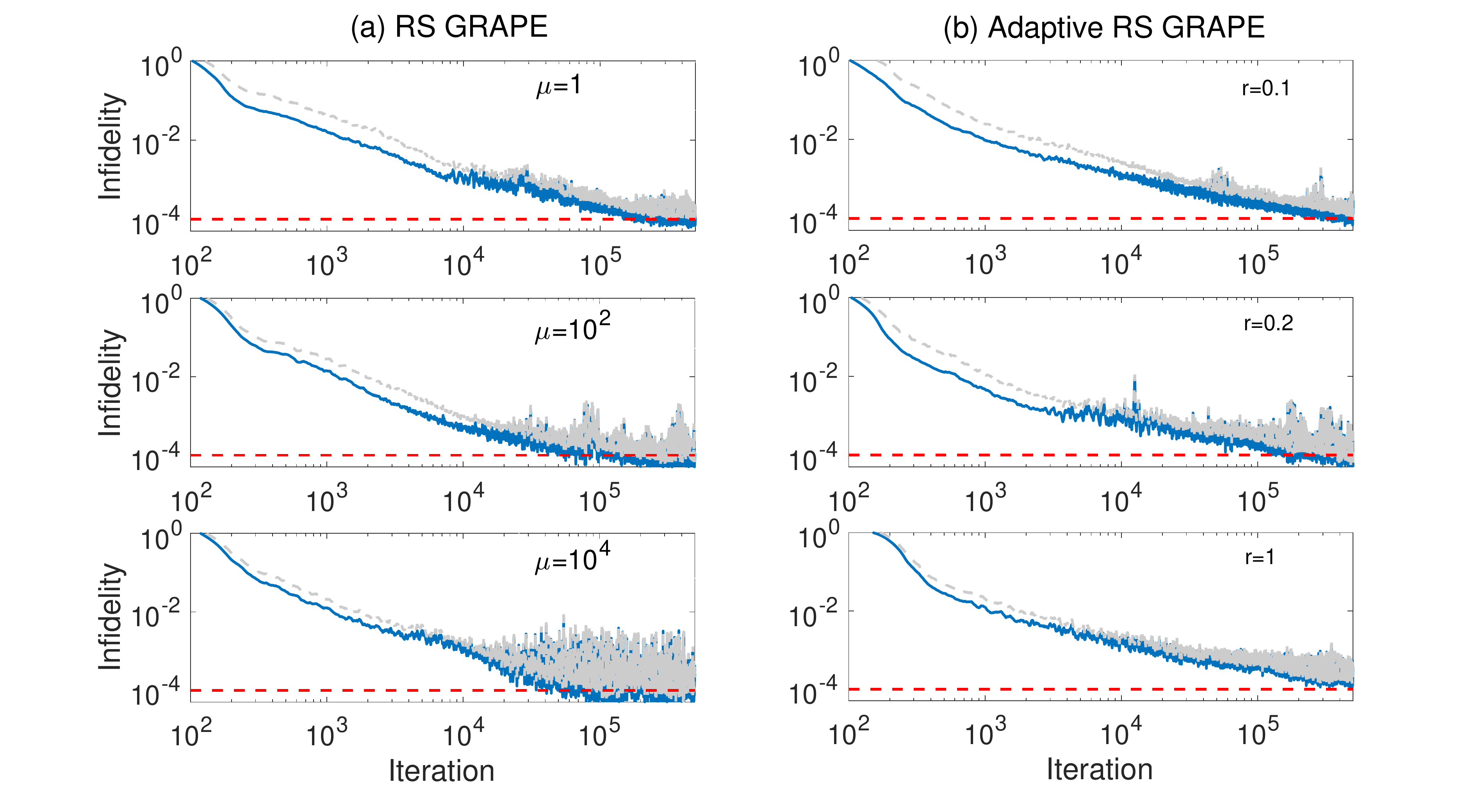}
	\end{center}
	\caption{The learning curves for the RS GRAPE approach ($1$st column) and the adaptive RS GRAPE approach ($2$nd column), respectively. The gray dashed line corresponds to the worst-case infidelity $J_{\text{max}}$ while the blue solid line corresponds to the average infidelity $J_{\text{mean}}$.}
	\label{learningcurve}
\end{figure*}

Based on the exponential utility, we first test the RS GRAPE algorithm with different sensitivity parameters $\mu=1,\ 10^2$ and $10^{4}$. Figure~\ref{learningcurve}(a) displays the resulting learning curves, namely, the achieved average infidelity $J_{\text{mean}}$ and the worst infidelity $J_{\text{max}}$ over the sampled uncertainties versus the number of iterations. As seen from the curves, the control robustness, quantified by the average or worst-case infidelity, is greatly enhanced during the training. The learning curves are relatively smooth when $\mu$ is small (e.g., $\mu=1$), but converge poorly when $\mu$ is large (e.g., $\mu=10^4$) because they oscillate drastically. 

\begin{figure}
	\begin{center}
		\includegraphics[width=0.9\columnwidth]{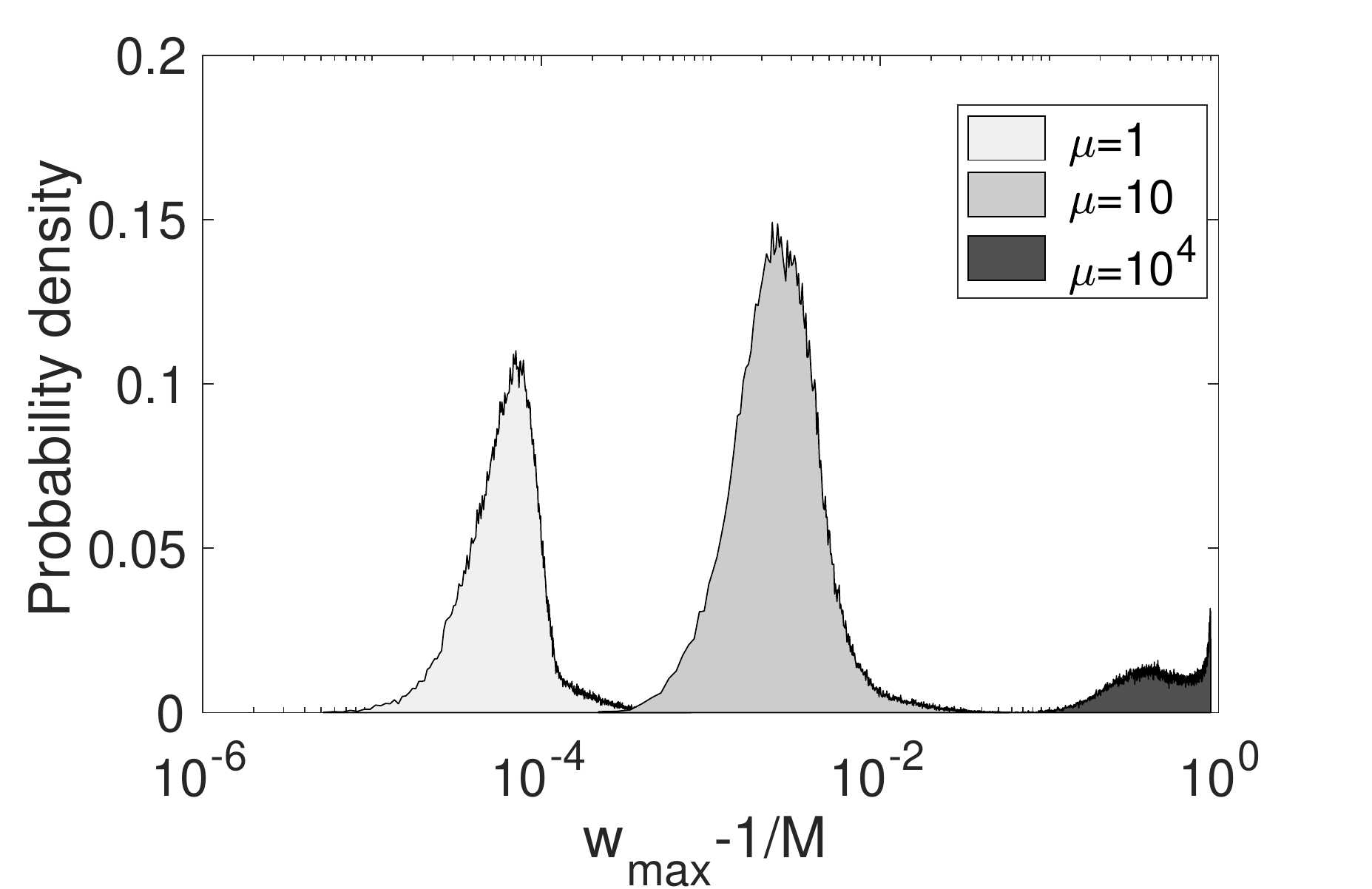}
	\end{center}
	\caption{The probability density function (pdf) versus the diversity degree (i.e., $\omega_{\text{max}}=\max_i\omega_i$) under the controls optimized by the RS GRAPE algorithm with $\mu=1,\ 10^2$ and $10^4$, respectively. The pdfs are estimated over $10^5$ uncertainty sample batches.}
	\label{Pm}
\end{figure}

As mentioned above, the diversity of uncertainty training samples may gradually decrease with the improved control robustness, which makes the training risk-insensitive. This can be observed when $\mu$ is not large for the exponential utility. For exmaple, in the case $\mu=1$, the average infidelity can be greatly suppressed to be below $10^{-3}$ after thousands of iterations, and thus $e^{\mu L}\approx 1$ for almost all uncertainty samples, i.e., the uncertainty samples will be equally weighted. This implies that the succeeding training is approximately subject to the average infidelity. To see this, we selected the controls optimized after $10^4$ iterations, and obtain the statistics of the index $d=\max_{1\leq k\leq M}\omega_k-1/M$ that indicates the performance diversity. Their probability density distributions displayed in Fig.~\ref{Pm} clearly show that in the cases $\mu=1$ and $10^2$ (in particular, the case $\mu=1$), $\omega_{\text{max}}$ is concentrated near the value $1/M$. By contrast, in the case $\mu=10^4$, $\omega_{\text{max}}$ is more diversely distributed in the range $[1/M,1]$, which leads to the oscillatory learning curve.

\begin{figure}
	\begin{center}
		\includegraphics[width=0.9\columnwidth]{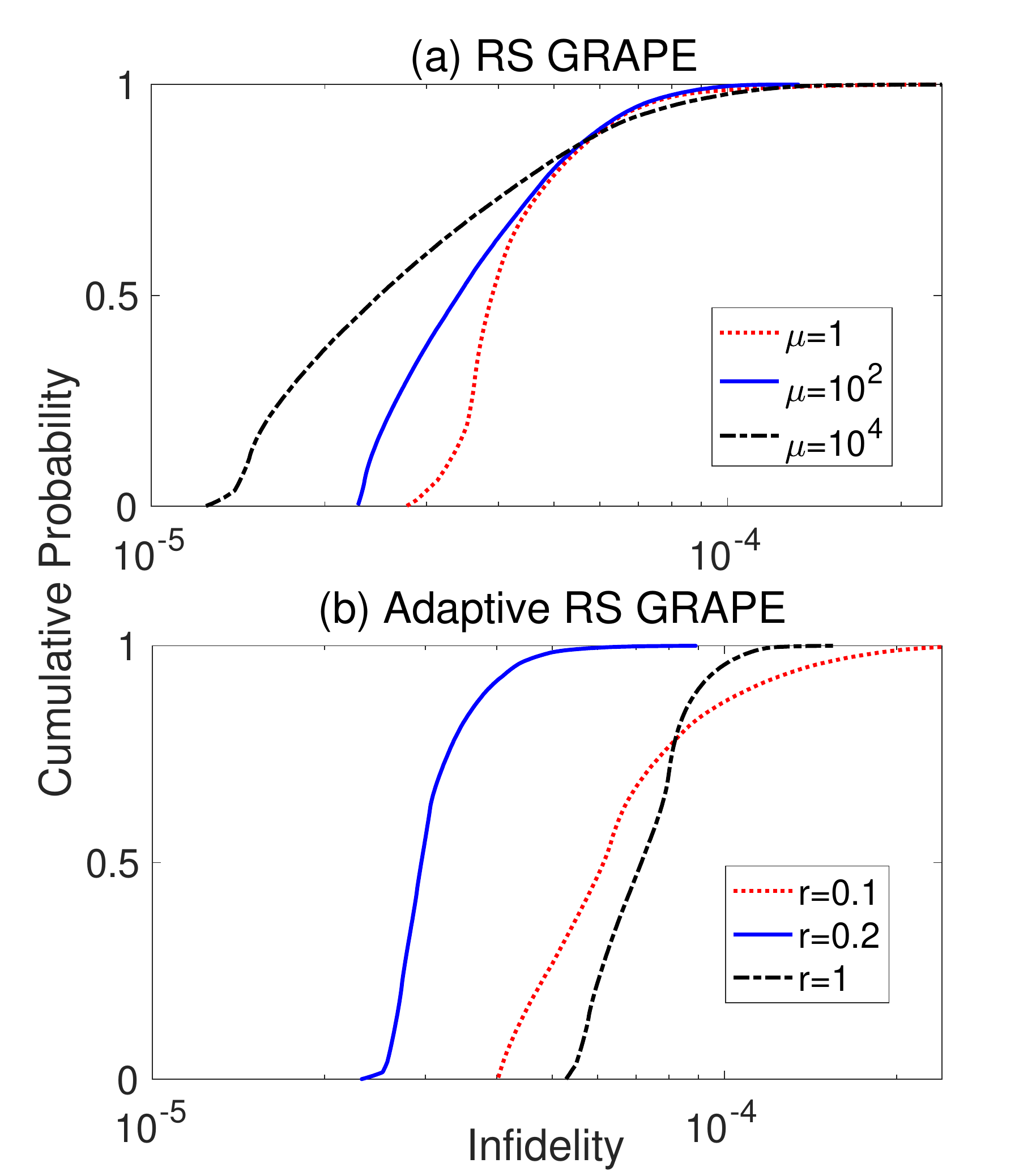}
	\end{center}
	\caption{The cumulative probability versus infidelity based on $10^5$ independent uncertainty samples under the controls optimized by the RS GRAPE approache and the adaptive RS GRAPE approach respectively.
	}
	\label{cdf}
\end{figure}

 To evaluate and compare the overall performance of the optimized control $\boldsymbol{u}_{\text{opt}}$, we can numerically calculate the cumulative distribution function (cdf) $F(l)$ of the gate infidelity, i.e., the probability for the infidelity $L(\boldsymbol{u}_{\text{opt}},\boldsymbol{\epsilon})$ being not larger than $l$. As shown in Fig.\ref{cdf}(a), the optimized control obtained in the case $\mu=10^4$ performs better in the relatively high-precision regime, but its worst-case infidelity is a little poorer due to the relatively instable training process.    

We also test the adaptive RS GRAPE approach with $r^*=0.1,\ 0.2$ and $1$, respectively. As seen in Fig.~\ref{learningcurve}(b), the robustness of the controls is rapidly enhanced by the training in all cases. We further display the cdfs under the optimized controls in Fig.~\ref{cdf}(b), from which we see that the case $r=0.2$ is overwhelmingly more robust as the entire cdf curve is above the other two, meaning that it performs better both in high-precision regime and worst-case infidelity. This demonstrates that the adaptive RS-GRAPE with an appropriate chosen diversity degree will lead to more robust controls.

\begin{figure}
 	\begin{center}
 		\includegraphics[width=0.9\columnwidth]{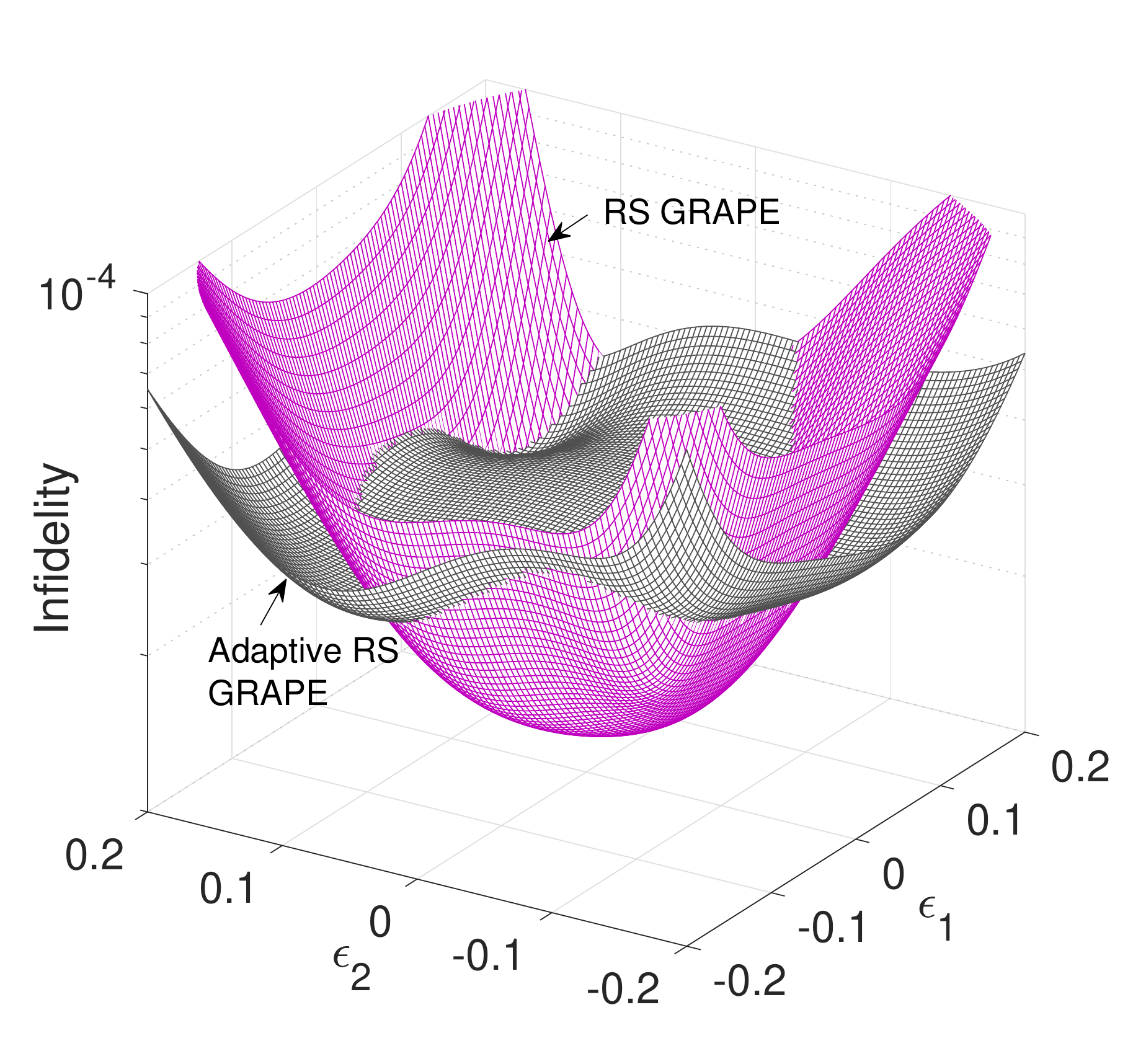}
 	\end{center}
 	\caption{ The infidelity versus two uncertainty parameters under controls optimized with the RS GRAPE approach ($\mu=10^4$) and the adaptive RS GRAPE approach ($r=0.2$), respectively.
 	}
 	\label{3D}
 \end{figure}

Compared with the RS GRAPE approach, the adaptive RS GRAPE approach performs more effectively in reducing the worst-case infidelity as the samples that yield poorest performance can be always heavily penalized. To better compare their overall performances, we plot in Fig.~\ref{3D} the $3$D landscapes of the infidelity as the function of the two uncertainty parameters under the optimized controls. The landscape associated with the adaptive RS GRAPE is relatively flat and is all below $10^{-4}$ in the displayed regime. The landscape associated with the RS GRAPE has high precision in the central part, but lower precision at the edges. This shows that the adaptive GRAPE achieves better worst-case performance, but its performance in the higher-precision regime is poorer. The control optimized by the RS GRAPE performs better when the uncertainty is relatively small. When the uncertainty varies in a larger regime, the control optimized by the adaptive RS GRAPE will be perferred.
 
In addition, it should be noted that the training processes of the adaptive RS GRAPE with $r=0.1$ and $r=1$ actually correspond to those subject to the average infidelity and the worst-case infidelity respectively. By comparison, the training based on the RS loss is much more effective.

\section{Conclusion}\label{discussion}

To conclude, we proposed quantum robust control design algorithms under the risk-sensitive (RS) criterion, which take advantages of the worst-case and average infidelities. The RS-GRAPE and adaptive RS-GRAPE algorithms are presented in which the sensitivity parameter is fixed and adaptively tuned, respectively. Numercial simulates demonstrate that these training algorithms can greatly enhance the control robustness, even when the uncertainties vary in a large regime. Compared with the existing a-GRAPE and b-GRAPE~\cite{Wu2018Deep} algorithms, the RS based GRAPE algorithms can remarkably improve the precision and robustness.

For practical applications, the proposed algorithms are advantageous in that only a few parameters (e.g., learning rate, batch size, etc.) are to be empirically tuned, as well as the sensitivity parameter and the utility function. The adaptive tuning strategy can also be flexibly chosen. In future studies, it is deserved to develop more effective strategies for updating these parameters so as to achieve stronger robustness.

\begin{acknowledgements}
	The author Re-Bing Wu acknowledges the support of the National Key R$\&$D Program of China (Grants No. 2018YFA0306703 and No. 2017YFA0304304) and NSFC (Grants No. 61833010 and No. 61773232).
\end{acknowledgements}

\bibliography{RS}

\end{document}